\documentclass[oldversion]{aa}
\usepackage{epsfig}
\usepackage{natbib}
\usepackage{lscape}
\usepackage{wasysym}
\usepackage{amssymb}
\usepackage{longtable}
\usepackage{rotating}
\usepackage{color}
\usepackage{colortbl}
\usepackage{fancyheadings}
\bibpunct{(}{)}{;}{a}{}{,}

\begin{document}

\title{Transit-timing measurements with the model-independent
barycenter method: Application to the LHS 6343 system}

\author{ M. Oshagh\inst{1,2} \and G. Bou{\'e}\inst{1} \and N. Haghighipour\inst{3}
\and M. Montalto\inst{1} \and P. Figueira \inst{1} \and N. C. Santos\inst{1,2} }

\institute{
Centro de Astrof{\'\i}sica, Universidade do Porto, Rua das Estrelas, 4150-762 Porto,
Portugal \\
email: {\tt moshagh@astro.up.pt}
\and
Departamento de F{\'i}sica e Astronomia, Faculdade de Ci{\^e}ncias, Universidade do
Porto,Rua do Campo Alegre, 4169-007 Porto, Portugal
\and
Institute for Astronomy and NASA Astrobiology Institute, University of Hawaii-Manoa,
2680 Woodlawn Drive, Honolulu, HI 96822,USA
}

\date{Received XXX; accepted XXX}

\abstract {We present a model-independent technique for calculating the time of mid-transits.
This technique, named ``barycenter method", uses the light-curve's symmetry to determine the
transit timing by calculating the transit light-curve barycenter. Unlike the other methods of
calculating mid-transit timing, this technique does not depend
on the parameters of the system and central star. We demonstrate the capabilities of the barycenter method
by applying this technique to some known transiting systems including several \emph{Kepler}
confirmed planets. Results indicate that for complete and symmetric transit lightcurves,
 the barycenter method achieves the same precision as other techniques,
but with fewer assumptions and much faster. Among the transiting systems studied with the barycenter method, we focus in particular on  LHS 6343C, a brown
dwarf that transits a member of an M+M binary system, LHS 6343AB. We present the
results of our analysis, which can be used to
set an upper limit on the period and mass of a possible second small perturber.}


\authorrunning{M. Oshagh et al.}
\titlerunning{Barycenter method and application.}
\maketitle

\section{Introduction}

The success of the transit-timing variation (TTV) method in characterizing planets around the stars Kepler 9
\citep{Holman-10} and Kepler 11 \citep{Lissauer-11} and in detecting a planet around star Kepler 19 \citep{Ballard-11}
strongly suggests that TTV method has come of age and is now among
the main mechanisms for detecting extrasolar planets.
This method which is based on modeling the variations that
appear in the times of the transits of a planet due to the perturbations of other objects,
has been shown by many authors to be capable of detecting small Earth-sized planets,
moons of giant planets, and stellar companions around variety of stars
\citep{Miralda-02,Holman-05,Agol-05,Kipping-09,Montalto-10,schwarz-11,HaghKirste-11}.

Because, the interaction between the transiting planet and the
perturbing body(ies) is gravitational if there are no magnetic fields, the amplitude of the TTV strongly depends on the masses of these objects
and their orbital architecture. As shown by
\citet{Holman-05}, \citet{Agol-05}, and \citet{HaghKirste-11},
the amplitude of a TTV signal varies with the mass and distance of the perturbing body.
The latter has been used in several null detections to place an upper limit on the mass and
orbital parameters of a hypothetical perturber
\citep{Bean-09,Csizmadia-10,Adams-10-1,Adams-11,Maciejewski-10,Maciejewski-11-1,Maciejewski-11-2}.

The TTV amplitude is strongly amplified when the transiting and perturbing planets are in a mean-motion
resonance.
For instance, as shown by \citet{Agol-05, Steffen-07, Agol-07,Haghighipour-09, HaghKirste-11},
a planet as small as Earth can produce large and detectable TTVs on a transiting
Jupiter-like body in or near a resonance.
This characteristic of resonant transiting systems makes the TTV method a powerful technique for
detecting low-mass planets.

The fact that different orbital configurations of the transiting and perturbing bodies can
produce similar TTVs has made the inference of the mass and orbital elements of the perturber
from the measurements of the transiting planet's TTVs a very complicated task. Several attempts
have been made to overcome
these difficulties \citep{Nesvorny-08, Nesvorny-09, Nesvorny-10, Meschiari-10}. However, the
complications still exist, particularly when the system is in or near a resonance. As shown by
\citet{Melendo-11}, continuous observations by Kepler and CoRoT are expected to resolve some of
these difficulties.

Determining variations in transit timing requires precise measurements of the times of mid-transits.
To compute a mid-transit time, it is necessary to develop a theoretical light-curve that best models the
observational measurements of the intensity of the light of a star. When studying transiting planets,
many authors use the analytical methodology
developed by \citet{Mandel-02} for this purpose. In this method, the light-curve of a star is calculated
using an analytical formula that contains several parameters such as the coefficients of
the star's limb darkening,
the ratio of the radius of the planet to that of the star, the semimajor axis of the planet (or its
orbital period), and the planet's orbital inclination. To measure the individual mid-transit times,
it is customary to hold all parameters (except mid-transit time) constant during the fitting procedure.
As a result, the measurement of the time of each mid-transit will be vulnerable to systematic
errors. In other words, any modification to the values of any of the above-mentioned parameters
(which may be obtained when observing the system for longer times) will change the fitted light-curve and
result in different values of the times of mid-transits. Subsequently, the values of the TTVs obtained in
these systems will also change.

We used a model-independent methodology, first introduced by \citet{Szabo-06}, to calculate the
time of mid-transit. We call this technique the {\it barycenter method} because it calculates the mid-transit
times by using the definition of the transit light-curve barycenter and its symmetry.
We describe this methodology in section 2 and present examples of its application to some of the already known
transiting systems in section 3. In section 4, we apply this technique to the system of LHS 6343 and explain
its implications for the transit timing of the system. We analyze the derived O-C diagram of LHS 6343
in section 5, and in section 6 we conclude this study by summarizing our analysis and reviewing the results.

\section{Barycenter method}

As mentioned earlier, to determine the variations in the transit timing of a planet,
a precise calculation of the times of its mid-transits is required. The mid-transit times
are determined by fitting an analytically obtained light-curve to the observational data,
and calculating the time of the mid-point of each individual transit on the latter curve.
When the transiting body is planetary, the synthetic light-curve is usually produced using the
algorithm developed by \citet{Mandel-02}. In the majority of cases, the times of mid-transits
are calculated by keeping all other parameters (e.g. stellar radius, planet radius, orbital period,
and two limb darkening coefficients) constant during the fitting procedure. This is particularly
important when the number of points inside a transit is small (e.g., 6-9 points). In such cases,
fitting the observed data can lead to imprecise results.

Another technique for calculating times of mid-transits is the ``trapezoid method" \citep{Alonso-09}.
In this method, a trapezoid function is fitted to the observational data and the best light-curve is
determined by varying the depth, duration, and shape of the trapezoid. The time of mid-transit is then
calculated by identifying the mid-point of each transit on the best-fit trapezoidal curve.

In systems where the transiting/eclipsing body is a stellar companion, the time of each  mid-transit/eclipse is
calculated using the methodology developed by \citet{kwee-56}. This method has been used by \citet{Deeg-00,Deeg-08}
to calculate eclipse timing variations of eclipsing binaries caused by a circumbinary planet, and
is based on the assumption that in an unperturbed system, the light-curve of the transited/eclipsed star is symmetric.
In this method, the mid-point of an eclipsing light-curve is determined by folding the light-curve around one point of the
transit, and calculating the differences between the points on the two parts of the folded light-curve. The point where these
differences become minimum corresponds to the point of mid-transit.

The method developed by \citet{kwee-56} has the advantage that unlike the method of \citet{Mandel-02}, it does not depend
on the parameters of the central star. However, for the measurements of the mid-transit times to be accurate,
this method requires very many points, which are obtained through
the interpolation of points from the results of observation.
As a result, in transiting systems with few data points (e.g., the transiting systems identified in
the long cadence of Q0 to Q2 data sets from {\it Kepler}), using this method is not practical.

In this section, we explain a methodology that employs similar idea as the method by \citet{kwee-56}
(i.e., using light-curve's symmetry) and as such is independent of the system's stellar parameters.
This method was first presented by \citet{Szabo-06} and later used by \citet{Simon-06} and \citet{Kipping-11} to study the possibility of the detection of exomoons. It calculates the exact
moment of mid-transit using the definition of the transit light-curve barycenter. We call this methodology the ``barycenter method".
Unlike the method by \citet{kwee-56}, the barycenter method can be applied to transit planetary systems with few
data points.

To define the transit light-curve barycenter, we use a normalized graph of the flux of the central star.
As shown in Figure 1, the flux of the star outside the transit is detrended and normalized to 1.
For a point $i$ with a flux $f_i$
inside the transit, the corresponding value of the light-loss of the system is equal to $1-{f_i}$. Similar to the barycenter
point of a number of massive objects, we now define a barycenter for the points on the graph of the normalized flux.
In this definition, the time of mid-transit will then be given by

\begin{equation}
T =\frac{\sum\limits_{i=1}^n t_{i}(1-f_{i})}{\sum\limits_{i=1}^n(1-f_{i})}.
\end{equation}

\noindent
In equation (1), \emph{n} represents the number (rank) of the data points in the observation, and $t_{i}$ is
Julian Day (JD) of the observation point $i$.
To obtain more precise results, we only consider the points that are inside the transit light-curve. A point is inside the transit if its light-loss $(1-fi)$ is higher than the standard deviation of flux outside the transit.

\begin{figure}
\center
\includegraphics[scale=0.38]{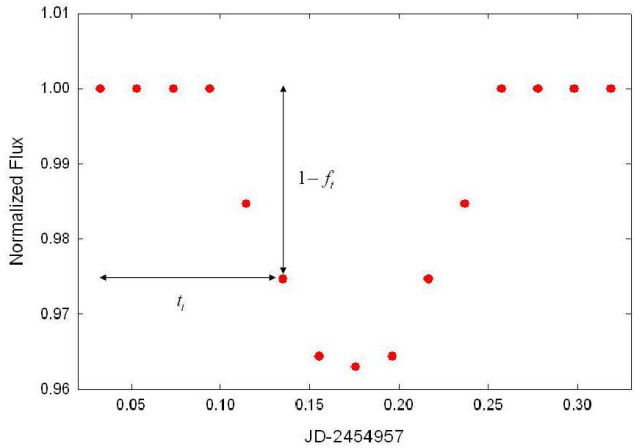}
\caption{Schematic view of the barycenter method.}
\label{fig:exslope}
\end{figure}

\section{Application of the barycenter method to known transiting systems}

\subsection{HAT-P-7b}
To test the capability of the barycenter method and the validity of its results, we used this technique
to calculate the times of mid-transits in several known transiting planetary systems.
In addition to the timing, transits may also
show variations in their durations and depths. However, our focus is only on the variations in the times of mid-transits.
Our first case was the transiting planet HAT-P-7b (Kepler-2b). HAT-P-7 was
observed in short and long cadences as a calibration target for \emph{Kepler}. In the short cadence mode,
the light-curve of HAT-P-7 consisted of approximately 355 points in each transit. We used the results of
the observations as reported in Q0 data set, and calculated the time of mid-transit for the first
transit of this planet. Figure 2 shows the results (the first point from the left). The error bar on each
point was determined using the bootstrap technique \citep{Wall-03}. As shown here,
the time of mid-transit obtained from the barycenter method is consistent with those obtained from
the trapezoid method and the model by \citet{Mandel-02}.

To evaluate the sensitivity of each of these techniques to the number of points in a transit,
we reduced the number of points in the light-curve by regular sampling, and calculated the time of mid-transit
using all three methods. Results are shown in Figure 2. As expected, the sizes of the error bars indicating
the uncertainties at each point increase for fewer data points. However, as Figure 2 shows,
the times of mid-transits obtained by all three methods are close and agree with one another.

\begin{figure}
\center
\includegraphics[scale=0.56]{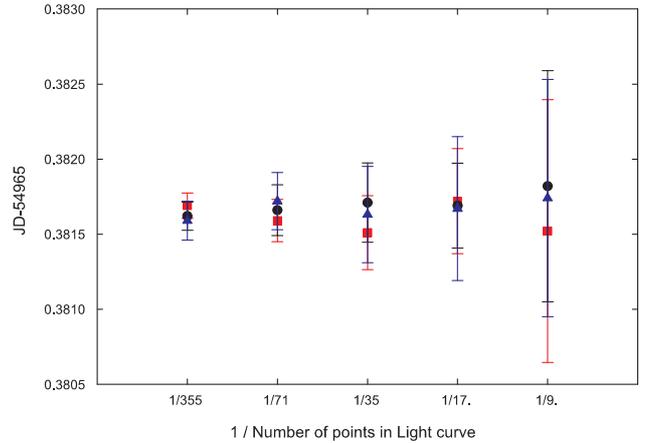}
\caption{Comparison between the results of mid-transit of HAT-P-7b first transit, obtained by
the Mandel \& Agol method (black circle), the trapezoid method (red square) and the barycenter method (blue triangle).}
\label{fig:exslope}
\end{figure}

\subsection{Kepler-1b to Kepler-9c}

We also applied the barycenter method to the confirmed planets of the Kepler-1 to Kepler-9 systems.
Table 1 shows the results and their corresponding uncertainties. The uncertainties were
calculated using the equation

\begin{equation}
  \sigma(T)^{2}=\sum\left|\frac{\partial T}{\partial f_{i}}\right|^{2} \sigma^{2}_{i}.
\label{eq:Lebseque I}
\end{equation}

\noindent
Table 1 also shows the values of the mid-transit times of these planets as reported by \citet{Holman-10} and \citet{Ford-11}
using the model of \citet{Mandel-02}. As shown here, the results obtained from the barycenter
method agree very well with the previously reported values.

Figures 3 and 4 show the differences between the values of mid-transit times obtained by the barycenter
method and those reported by \citet{Holman-10} and \citet{Ford-11}. The error bar at each point was calculated
by taking the quadratic sum of the uncertainties shown in Table 1. Table 2 lists standard deviations of these
differences and their average error bars. As can be seen from Figure 3 and Table 2, Kepler-4b shows large error bars
compared to those of other planets because of its shallow transits. This figure also shows that the standard
deviation of Kepler-3b, as listed in Table 2, is larger than its average error, which can be attributed to the non-symmetric
shapes of the first and sixth transits of this planet (see Fig.5). These short-lived anomaly flux variations can be explained
by different mechanisms such as the presence of active regions (dark spots) or a second transiting body
\citep{Rabus-09,Sanchis-Ojeda-11,Sanchis-Ojeda-11b,Silva-08,Nutzman-11,Deming-11}.

Since in both the barycenter method and the method of \citet{Mandel-02} it is assumed that the light-curve is symmetric,
these methods are sensitive to missing points in the observation of a transit. This can be seen from Table 2 for
Kepler-6b and Kepler-9b. The standard deviations of these two planets are larger than their average errors, which could
have been caused by a missing point in the observation of the first transit of Kepler-6b and fifth transit of Kepler-9b
(Fig. 6). To illustrate this effect, we made an artificial light-curve for a transiting planet and calculated the time of
its mid-transit using the model of \citet{Mandel-02}. We then removed one point from the light-curve and calculated
the time of mid-transit using both the barycenter method and the method of \citet{Mandel-02}.
As shown in Figure 7,  and in agreement with \citet{Csizmadia-10},
the results obtained by both methods show large deviations when the missing point was in ingress or
egress (deviation $\sim500$ seconds). On the other hand, both methods become less sensitive when the missing point is
close to the bottom of the light curve. This experiment suggested that both the barycenter and the Mandel \& Agol methods
require the full coverage of observation data in transit, and a missing point in the observation data may cause a large
offset in the results.

\begin{figure}[t]
\center
\includegraphics[scale=0.55]{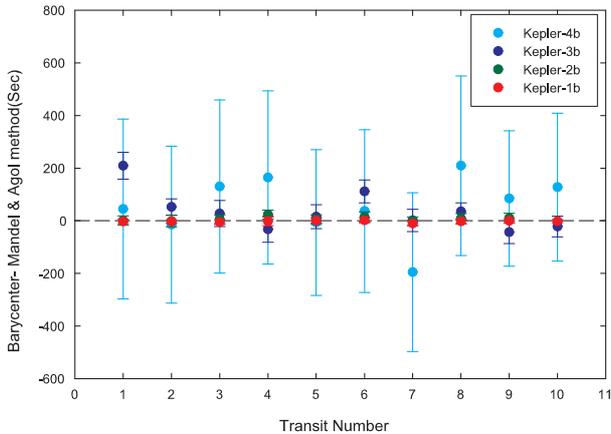}
\caption{Differences between the values of mid-transit timing obtained by the barycenter method and
the Mandel \& Agol method for Kepler-1b to Kepler-4b \citep{Holman-10, Ford-11}.}
\label{fig:exslope}
\end{figure}

\begin{figure}[t]
\center
\includegraphics[scale=0.55]{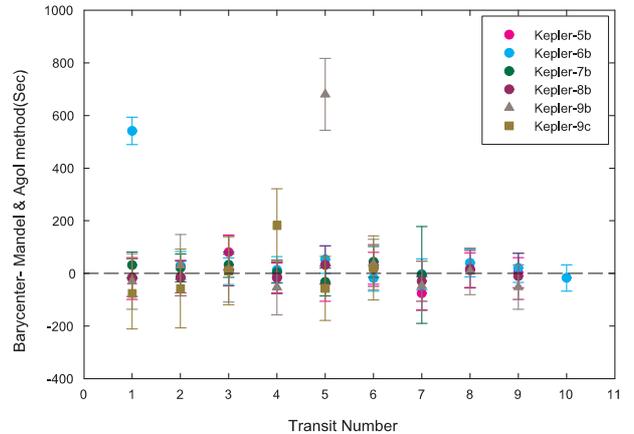}
\caption{Same as in Fig.3 for Kepler-5b to Kepler-9c.}
\label{fig:exslope}
\end{figure}

We also examined the applicability of the barycenter method to the long integration time of \emph{Kepler}'s long cadence
observations (29.42 minutes) \citep{Kipping-10}.  Using the algorithm by \citet{Mandel-02}, we generated an artificial
light-curve with bins of long integration times (we chose a point every 6 seconds and used the mean of 270 of those
points as the observed flux). We changed the beginning time of each binning and studied the variations of mid-transit
times as determined by the barycenter method. Results point to a deviation of no more than 4 seconds for the mid-transit times.

\begin{table}
\tiny
\caption{Transit timing of Kepler-1b - Kepler-9c, measured by both methods, the first column is the number of
transit, the second column lists mid-transits found by the barycenter method(MJD-2454900), and the third column lists mid-transits
reported in \citet{Holman-10, Ford-11}.}

\begin{center}
\begin{tabular}{c c c c c c c}

\hline
	Number & ${T_0}$-Barycenter method(MJD-2454900)  &${T_0}$(MJD-2454900) \\
\hline

1.1 & 65.645006$\pm$0.000054 & 65.645036$\pm$0.000073 \\
1.2 & 68.115621$\pm$0.000051 & 68.115649$\pm$0.000051 \\
1.3 & 70.586201$\pm$0.000051 & 70.586262$\pm$0.000079 \\
1.4 & 73.056852$\pm$0.000055 & 73.056875$\pm$0.000186 \\
1.5 & 75.527497$\pm$0.000051 & 75.527488$\pm$0.000098 \\
1.6 & 77.998120$\pm$0.000054 & 77.998101$\pm$0.000103 \\
1.7 & 80.468596$\pm$0.000052 & 80.468714$\pm$0.000071 \\
1.8 & 82.939301$\pm$0.000051 & 82.939327$\pm$0.000080 \\
1.9 & 85.409937$\pm$0.000050 & 85.409940$\pm$0.000074 \\
1.10 & 87.880533$\pm$0.000051 & 87.880553$\pm$0.000054 \\
\hline
2.1 & 65.381504$\pm$0.000085 & 65.381496$\pm$0.000185 \\
2.2 & 67.586213$\pm$0.000087 & 67.586232$\pm$0.000238 \\
2.3 & 69.791026$\pm$0.000090 & 69.790967$\pm$0.000178 \\
2.4 & 71.995973$\pm$0.000088 & 71.995703$\pm$0.000178 \\
2.5 & 74.200488$\pm$0.000113 & 74.200438$\pm$0.000181 \\
2.7 & 78.610067$\pm$0.000089 & 78.609909$\pm$0.000210 \\
2.8 & 80.814614$\pm$0.000086 & 80.814644$\pm$0.000172 \\
2.9 & 83.019473$\pm$0.000086 & 83.019380$\pm$0.000210 \\
2.11 & 87.428970$\pm$0.000086 & 87.428851$\pm$0.000199 \\
\hline
3.1 & 67.591623$\pm$0.000181 & 67.587958$\pm$0.000558 \\
3.2 & 72.476370$\pm$0.000096 & 72.475763$\pm$0.000343 \\
3.3 & 77.363887$\pm$0.000111 & 77.363568$\pm$0.000564 \\
3.4 & 82.250995$\pm$0.000114 & 82.251373$\pm$0.000555 \\
3.5 & 87.139357$\pm$0.000103 & 87.139178$\pm$0.000515 \\
3.6 & 92.026855$\pm$0.000108 & 92.026983$\pm$0.000491 \\
3.7 & 96.914806$\pm$0.000094 & 96.914788$\pm$0.000482 \\
3.9 & 106.690806$\pm$0.000096 & 106.690398$\pm$0.000368 \\
3.10 & 111.577701$\pm$0.000096 & 111.578203$\pm$0.000488 \\
3.11 & 121.353559$\pm$0.000097 & 121.353813$\pm$0.000448 \\
\hline
4.1 & 104.816415$\pm$0.002636 & 104.815900$\pm$0.002940 \\
4.2 & 108.029416$\pm$0.002417 & 108.029584$\pm$0.002460 \\
4.3 & 111.244774$\pm$0.002342 & 111.243268$\pm$0.003001 \\
4.5 & 117.672537$\pm$0.002587 & 117.670635$\pm$0.002792 \\
4.6 & 120.884243$\pm$0.002438 & 120.884319$\pm$0.002080 \\
4.7 & 124.098430$\pm$0.002747 & 124.098003$\pm$0.002296 \\
4.8 & 127.309430$\pm$0.002302 & 127.311687$\pm$0.002627 \\
4.9 & 130.527792$\pm$0.002522 & 130.525370$\pm$0.003034 \\
4.11 & 136.953717$\pm$0.001792 & 136.952738$\pm$0.002374 \\
4.12 & 140.167899$\pm$0.001699 & 140.166422$\pm$0.002767 \\
\hline
5.1 & 66.545846$\pm$0.000776 & 66.546068$\pm$0.000494 \\
5.2 & 70.094334$\pm$0.000582 & 70.094536$\pm$0.000520 \\
5.3 & 73.643932$\pm$0.000517 & 73.643005$\pm$0.000564 \\
5.4 & 77.191271$\pm$0.000503 & 77.191473$\pm$0.000434 \\
5.5 & 80.739494$\pm$0.000514 & 80.739942$\pm$0.000579 \\
5.6 & 84.288634$\pm$0.000449 & 84.288410$\pm$0.000543 \\
5.7 & 87.836005$\pm$0.000510 & 87.836878$\pm$0.000526 \\
5.8 & 91.385478$\pm$0.000587 & 91.385347$\pm$0.000495 \\
5.9 & 94.933849$\pm$0.000492 & 94.933815$\pm$0.000447 \\
\hline
6.1 & 67.430812$\pm$0.000408 & 67.424550$\pm$0.000441 \\
6.2 & 70.659594$\pm$0.000359 & 70.659250$\pm$0.000499 \\
6.3 & 73.894048$\pm$0.000406 & 73.893951$\pm$0.000421 \\
6.4 & 77.128802$\pm$0.000352 & 77.128650$\pm$0.000468 \\
6.5 & 80.363963$\pm$0.000403 & 80.363351$\pm$0.000453 \\
6.6 & 83.597843$\pm$0.000355 & 83.598052$\pm$0.000443 \\
6.7 & 86.832717$\pm$0.000359 & 86.832752$\pm$0.000568 \\
6.8 & 90.067911$\pm$0.000355 & 90.067452$\pm$0.000489 \\
6.9 & 93.302393$\pm$0.000410 & 93.302153$\pm$0.000495 \\
6.10 & 96.536655$\pm$0.000356 & 96.536853$\pm$0.000456 \\
\hline
7.1 & 67.276389$\pm$0.000429 & 67.276027$\pm$0.000374 \\
7.2 & 72.161758$\pm$0.000406 & 72.161517$\pm$0.000459 \\
7.3 & 77.047387$\pm$0.000407 & 77.047008$\pm$0.000382 \\
7.4 & 81.932553$\pm$0.000332 & 81.932498$\pm$0.000316 \\
7.5 & 86.817611$\pm$0.000410 & 86.817988$\pm$0.000455 \\
7.6 & 91.703986$\pm$0.000412 & 91.703478$\pm$0.000538 \\
7.7 & 96.588905$\pm$0.000367 & 96.588969$\pm$0.002094 \\
\hline
8.1 & 64.685860$\pm$0.000691 & 64.686046$\pm$0.000472 \\
8.2 & 68.208387$\pm$0.000598 & 68.208545$\pm$0.000345 \\
8.3 & 71.731198$\pm$0.000581 & 71.731044$\pm$0.000385 \\
8.4 & 75.253391$\pm$0.000578 & 75.253544$\pm$0.000447 \\
8.5 & 78.776414$\pm$0.000555 & 78.776043$\pm$0.000628 \\
8.7 & 85.821393$\pm$0.000695 & 85.821041$\pm$0.000602 \\
8.8 & 89.343196$\pm$0.000618 & 89.343540$\pm$0.000630 \\
8.9 & 92.866237$\pm$0.000492 & 92.866039$\pm$0.000644 \\
8.10 & 96.388414$\pm$0.000731 & 96.388538$\pm$0.000712 \\
\hline
9b.1 & 77.2484$\pm$0.00086 & 77.24875$\pm$0.00087 \\
9b.2 & 96.48276$\pm$0.00099 & 96.4824$\pm$0.00092 \\
9b.3 & 134.95455$\pm$0.00122 & 134.95437$\pm$0.00077 \\
9b.4 & 154.18997$\pm$0.00094 & 154.19058$\pm$0.00077 \\
9b.5 & 173.44199$\pm$0.00116 & 173.43412$\pm$0.00107 \\
9b.6 & 211.92629$\pm$0.00082 & 211.92589$\pm$0.00074 \\
9b.7 & 231.17112$\pm$0.00082 & 231.17167$\pm$0.00071 \\
9b.8 & 250.42960$\pm$0.00074 & 250.42951$\pm$0.00071 \\
9b.9 & 269.68043$\pm$0.00071 & 269.68103$\pm$0.00068 \\
\hline
9c.1 & 69.30489$\pm$0.00091 & 69.30577$\pm$0.00127 \\
9c.2 & 108.33019$\pm$0.00133 & 108.33086$\pm$0.00111 \\
9c.3 & 147.33572$\pm$0.00107 & 147.3356$\pm$0.00105 \\
9c.4 & 186.31434$\pm$0.00120 & 186.31251$\pm$0.00107 \\
9c.5 & 225.26218$\pm$0.00103 & 225.26284$\pm$0.00096 \\
9c.6 & 264.18192$\pm$0.00099 & 264.18168$\pm$0.00100 \\
\hline
\end{tabular}
\end{center}
\label{default}
\end{table}

\begin{table}[htdp]
\caption{Comparison of standard deviation of the difference between
two methods and the average error bar obtained by quadratic sum.}
\begin{center}
\begin{tabular}{lcc}

\hline
Planet & Standard deviation (sec)& Average error (sec)\\
\hline
Kepler-1b & 3.3 & 8.9 \\
Kepler-2b & 8.2 & 18.6\\
Kepler-3b & 76.0 & 42.7 \\
Kepler-4b & 115.0 & 306.3 \\
Kepler-5b & 43.0 & 65.0 \\
Kepler-6b & 168.2 & 52.3\\
Kepler-7b & 26.5 & 69.3\\
Kepler-8b & 22.4 & 71.3\\
Kepler-9b & 233.2 & 105.7\\
Kepler-9c & 87.0 & 132.7\\
\hline
\end{tabular}
\end{center}
\label{default}
\end{table}%

\begin{table}[htdp]
\caption{Parameters of the LHS 6343 system according to \citet{Johnson-11}.}
\begin{center}
\begin{tabular}{lll}
\hline
Parameter  &$\>\>\>$& Value  \\
\hline
$M_{\rm A}$ (solar mass) &$\>\>\>$& 0.370 $\pm$ 0.009 \\
$M_{\rm B}$ (solar mass)  &$\>\>\>$& 0.30 $\pm$ 0.01 \\
$M_{\rm C}$ (Jupiter mass)  &$\>\>\>$& 62.7 $\pm$ 2.4 \\
$a_{\rm AB}$ (AU) &$\>\>\>$& 20.130 $\pm$ 0.605\\
$a_{\rm AC}$ (AU) &$\>\>\>$& 0.0804 $\pm$ 0.0006\\
$P_{\rm C}$ (days) &$\>\>\>$& 12.71382 $\pm$ 0.00004\\
\hline
\end{tabular}
\end{center}
\label{default}
\end{table}%

\section{The LHS 6343 system and its transit timing }

LHS 6343 is a close, M+M binary system with a separation of $\sim 20$ AU. The primary of this binary, LHS 6343 A
(KIC 10002261, RA=19h 10m 14.33s, Dec= $46\,^{\circ}$ 57$'$ 25.50$''$), has a mass of 0.37 $M_{\odot}$ and the mass of the secondary,
LHS 6343 B, is approximately 0.30 $M_{\odot}$ (see Table 3). The publicly available Q0 and Q1 data sets from {\it Kepler}
revealed four deep transits in the light-curve of this system. By analyzing these data, \citet{Johnson-11} showed that these transits are produced by
a third object, LHS 6343 C, which orbits LHS 6343 A every 12.71 days. As determined by these authors, LHS 6343 C
is a brown dwarf with a mass of $\sim 63 {M_{\rm J}}$ and is located at a distance 0.08 AU from LHS 6343 A.

In preparation for applying the barycenter method to the light-curve of LHS 6343, we analyzed each transit
of this system separately.
Our initial analysis of the light-curve of LHS 6343 at the time of the release of Q0 and Q1 data sets pointed
to a non-symmetric transit (transit number 3) among the initial four transits of this system.
We recall that the barycenter method is based on the symmetry of the shape of a transit.
The release of the Q2 data set provided us with seven more transits, of which our analysis identified transits number 5 and 8
as asymmetric. Figure 8 shows these non-symmetric transits.
We note that these anomalies may be caused by starspots.
To better portray the anomalies in the shapes of
these transits, we used Mandel \& Agol's methodology and obtained the best fit to all 11 transits of this system.
Figure 9 shows the residuals of each single transit with respect to this best fit.
As shown here, the residuals of the third, fifth, and eighth transits are larger than $1-\sigma$ (closer to $2-\sigma$)
because of their anomalies inside their transits.

As mentioned for Kepler-9b,
developing a model to explain these anomalies would require many observational points inside each transit and
will depend on several parameters such as the size and latitude of starspots,
their lifetimes, the rotational period of the star, and the orientation of the rotation axis of the star relative to the
orbit of transiting brown dwarf. Although an interesting project on its own, developing such a model is beyond the scope
of our study. Also, given that with the currently available data, the number of points in each transit
is limited to only 5 or 6, such a model may not even be entirely realistic.
Therefore, because the basis of the barycenter method is on the symmetry of a transit, and also to
restrain false positive TTVs, we decided to exclude the most asymmetric transits (i.e., transits 3, 5, and 8)
from our analysis. Table 4 lists the times of the mid-transits of the remaining eight transits of the system
calculated using the barycenter method. To estimate the corresponding errors of each mid-transit time,
we used the bootstrap method \citep{Wall-03} and considered the standard deviation inside each transit as the
initial uncertainty.
Note that the standard deviation inside a transit may be larger than outside due to crossing over starspots.
To check the validity of our error estimation, we also calculated the values of errors using equation (2)
and the methodology used by \citet{Doyle-04}. Our calculations showed that the values of the errors obtained
from all three methods have the same order of magnitude.

\begin{figure}[t]
\center
\includegraphics[scale=0.58]{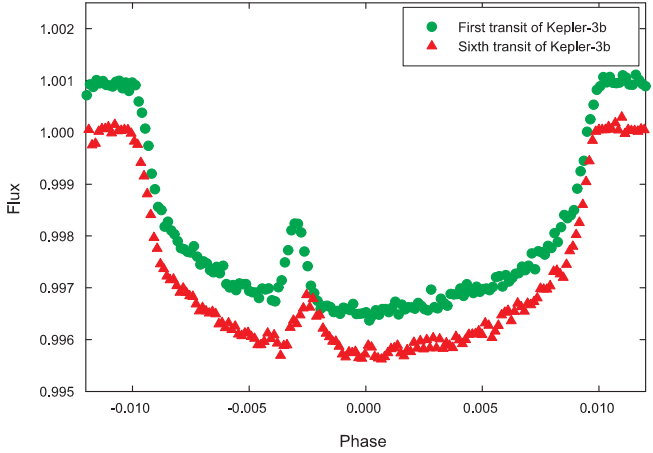}
\caption{Asymmetries in the first and sixth transits of Kepler-3b
 (maybe due to starspots).}
\label{fig:exslope}
\end{figure}

\begin{figure}[t]
\center
\includegraphics[scale=0.58]{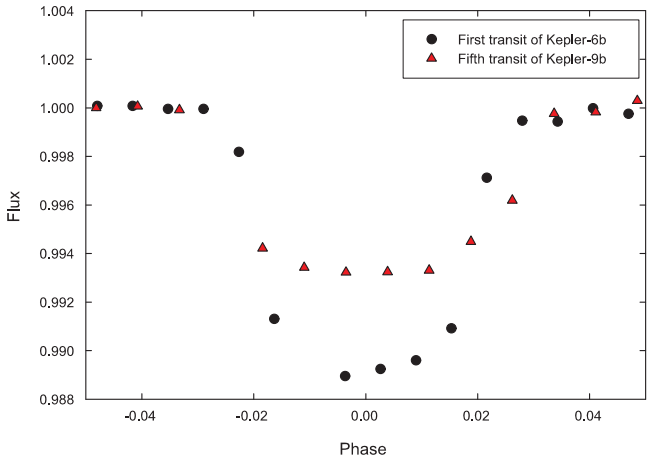}
\caption{Asymmetries in the first transit of Kepler-6b and fifth transit of Kepler-9b,
caused by missing point of observation.}
\label{fig:exslope}
\end{figure}

\begin{figure}[t]
\center
\includegraphics[scale=0.58]{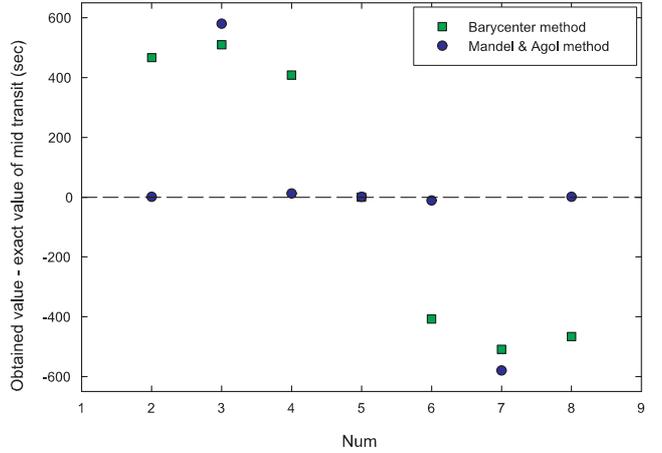}
\caption{Deviations of mid-transit timing from the known values as calculated by Mandel \& Agol, and
the barycenter methods
for a synthetic light curve with one missing point. The $x$-axis presents the rank of missing points in light curve.}
\label{fig:exslope}
\end{figure}

\begin{table*}[htdp]\scriptsize

\caption{Transit timing of LHS 6343, as measured by the barycenter method and O-C values in days were
calculated according to the new linear ephemeris.}
\begin{center}
\begin{tabular}{c c c c c c c}

\hline
Transit & $T_0$ (days)  & $T_0$ (days)  & O-C (Sec)  \\
Number  & (The barycenter Method) & (Calculated) & (TTV)  \\
\hline
1 & 54957.216473$\pm$0.000133 & 54957.216535& -5.4$\pm$11.5 \\
2 & 54969.930434$\pm$0.000154 & 54969.930354& 6.9$\pm$13.3\\
4 & 54995.358025$\pm$0.000135 & 54995.357992& 2.9$\pm$11.7\\
6 & 55020.785698$\pm$0.000120 & 55020.785630& 5.9$\pm$10.4\\
7 & 55033.499199$\pm$0.000147 & 55033.499449& -21.6$\pm$12.7\\
9 & 55058.927144$\pm$0.000136 & 55058.927087& 4.9$\pm$11.8\\
10 & 55071.641056$\pm$0.000173 & 55071.640906& 13.0$\pm$15.0\\
11 & 55084.354626$\pm$0.000156 & 55084.354725& -8.6$\pm$13.5\\
\hline
\end{tabular}
\end{center}
\label{default}
\end{table*}%

To obtain the variations in the transit timing of the system, we applied a linear fit to the eight
mid-transit times in Table 4. Results suggested a period of $P=12.713815$ days,
corresponding to a semimajor axis of 0.076-0.080 AU
for the transiting body. These results closely agree with the results reported by \citet{Johnson-11}.

Given that LHS 6343 is a binary system
and the transiting object (LHS 6343 C) orbits the primary star, the three-body system of LHS 6343 AC-B forms a hierarchical three-body system. We examined
the stability of LHS 6343 C in this system by numerically integrating its orbit.
Results indicated that this object is stable for long times. We refer the reader to a recent
article by \citet{Borkovits-11} and references therein, where the authors have presented a detailed
analysis of the dynamics and transit/eclipse timing variations of hierarchical tripe systems.

\begin{figure}[t]
\center
\includegraphics[scale=0.6]{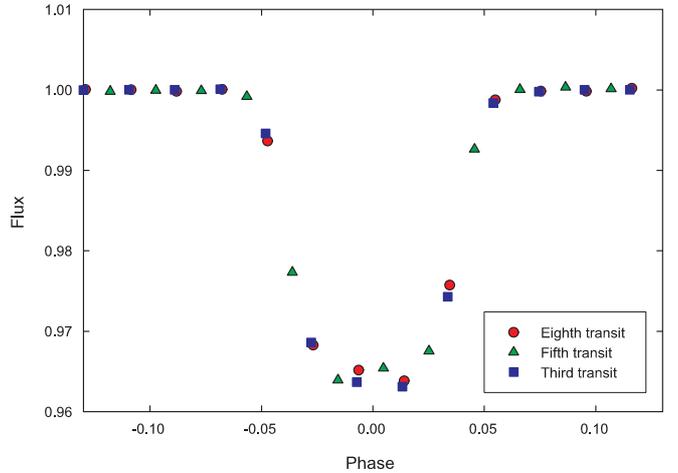}
\caption{Strange non-symmetric shape of third (blue square), fifth (green triangle), and eighth
(red circle) transit of LHS 6343. }
\label{fig:exslope}
\end{figure}

\begin{figure}[t]
\center
\includegraphics[scale=0.58]{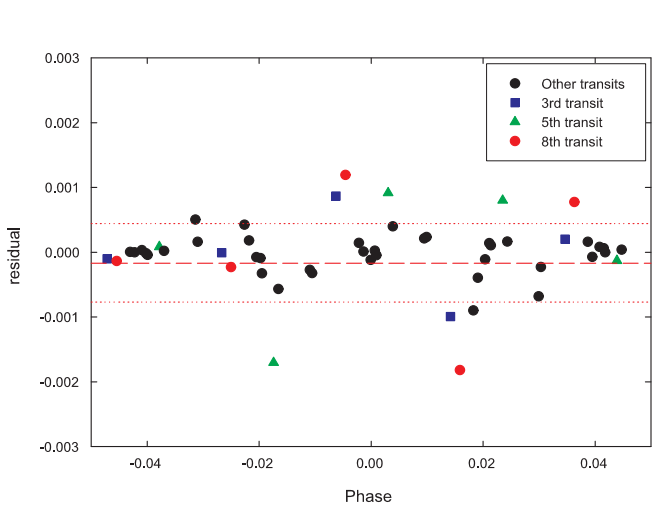}
\caption{Residual of the best fit of the Mandel \& Agol method to all transits of
LHS 6343 (just inside transit).}
\label{fig:exslope}
\end{figure}

\begin{figure}[t]
\center
\includegraphics[scale=0.58]{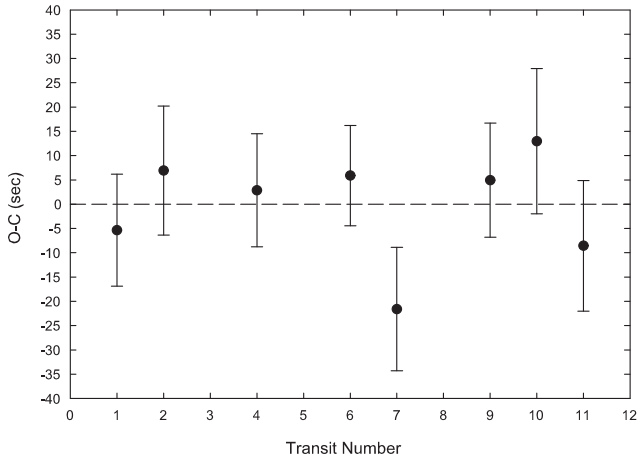}
\caption{Diagram of the transit-timing variations of LHS 6343.}
\label{fig:exslope}
\end{figure}

\begin{figure}
\includegraphics[width=8.5cm]{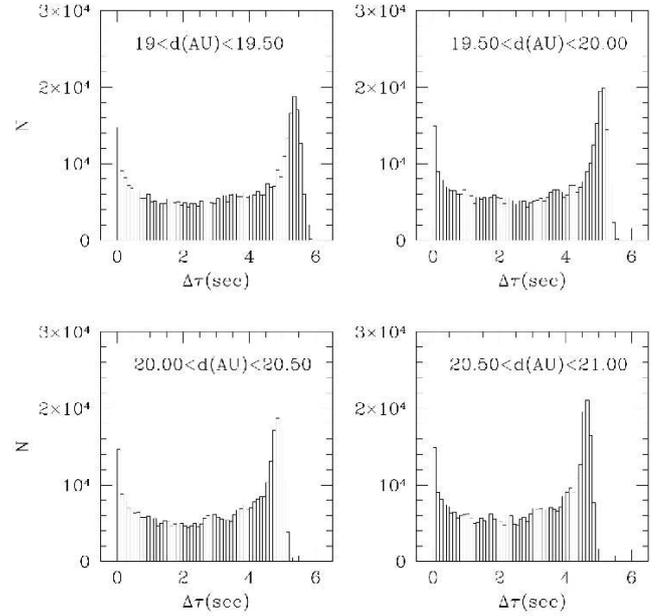}
\caption{Values of LTTs of LHS 6343 for different values of the binary semimajor axis.}
\label{fig11}
\end{figure}

\begin{figure}
\includegraphics[width=8.5cm]{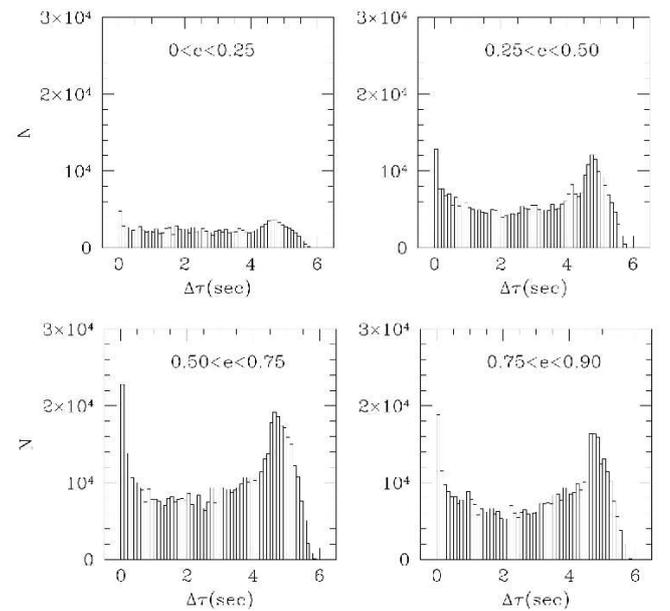}
\caption{Values of LTTs of LHS 6343 for different values of the binary eccentricity.}
\label{fig12}
\end{figure}

\section{Analyzing the O-C diagram of LHS 6343}

The times of mid-transits obtained from the barycenter method show small deviations
from their linear fit. Figure 10 and the right column of Table 4 show these deviations
and their corresponding uncertainties for each mid-transit time.
In this section, we analyze these deviations from the linear fit of transit timings and discuss their implications for the possible
existence of a second smaller object around the primary LHS 6343 A.

Because it is in a hierarchical tripe configuration, LHS 6343 C is continuously
subject to the gravitational perturbation of the secondary star. These perturbations
affect the orbit of this object and cause variations in the times of its transit
(for a detailed analysis of TTVs in hierarchical tripe systems we refer the reader
to Borkovits et al. 2011).
Given that the semimajor axis of the binary ($\sim$ 20 AU) and its projected separation
(19-21 AU) are much larger than the semimajor axis of
LHS 6343 C, it would be important to determine to what degree the variations in the transit
timing of this object have been caused by the binary's light-travel time (LTT) effect.
To examine this possibility, we used the methodology presented by
\citet{Montalto-10} and calculated LTTs for different values of the semimajor axis and eccentricity
of the binary. We changed the values of the projected separation of the binary using the distribution
given by \citet{Duquennoy-91}, and performed 10000 LTT-calculations for randomly chosen values of the
binary eccentricity between 0 and 0.9.
In all our simulations, we considered the system to be coplanar. We identified the systems for which the value of
LTT was between 1 s and 6 s. Figures 11 and 12 show the results for a timespan of three years (duration of {\it Kepler}'s
primary mission). As shown here, systems with LTTs between 4.5 s and 5.5 s constitute the majority of the cases
(we recall that the minimum reported value of TTV that can be detected by {\it Kepler} is $\sim$10 s see
\citep{Ford-11}).
Figure 13 shows the results of all our simulations for LTTs between 5 s and 6 s in more detail. As shown in this figure,
the values of LTTs do not exceed 6 s, which implies that during the period spanned by the present public
release of the \emph{Kepler} observations data ($\sim$144 days), the contribution of the  binary
LTT to the variations in the transit timing of LHS 6343 C is negligible.

\begin{figure}
\includegraphics[width=8.5cm]{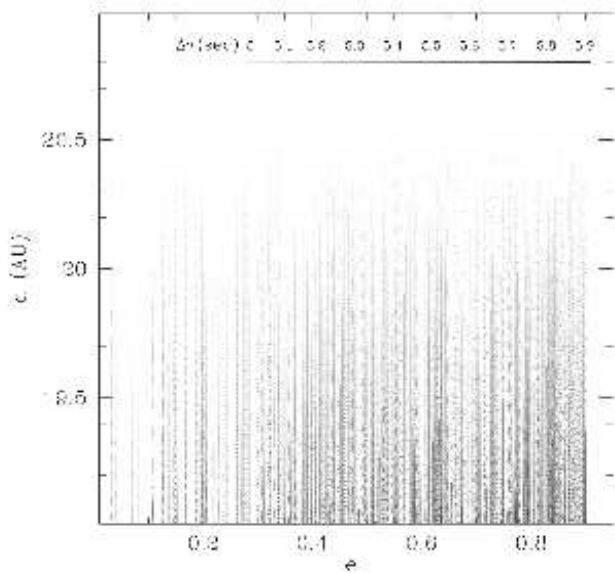}
\caption{Graph of LTTs between 5 s and 6 s for different values of the binary semimajor axis and eccentricity.}
\label{fig13}
\end{figure}

The fact that the contribution of LTT to the O-C values, as listed in
Table 4, is negligibly small implies that in modeling these deviations from the linear fit of transit timings, one can safely ignore the
effect of the secondary star.

To examine whether these deviations could be caused by an additional body
in the system, we considered the two-body system of LHS 6343 AC, and calculated the O-C values for different values of the mass, semimajor axis, and orbital eccentricity of a hypothetical
perturber around the primary LHS 6343 A. To reduce the amount of calculations (which could be large because of the large size
of the parameter-space), we limited our study to only circular and coplanar systems.
Figure 14 shows the results for different values of the initial angular position of the
hypothetical third body. A comparison between these results and the values of the O-C in the
right column of Table 4 suggests that a perturber with a mass ranging from 0.1
to 1 $M_{\rm J}$ may be able to produce these values when in an orbit with a period
ranging from $\sim 3.5$ to $8 {P}$ (where $P=12.713815$ days is the orbital period of LHS 6343 C)
around LHS 6343 A. To determine an upper limit for the mass of the perturber, we calculated the O-C values
for different values of the mass and semimajor axis of this object, and compared the results
with the values of O-C as shown in Figure 10. Figure 15 shows the maximum values
of the mass of the perturber for which the value of $\chi^2$ between the O-C obtained from the model
and those listed in Table 4 are lower than 3. As shown in this figure, the mass of the perturber cannot be larger
than one Jupiter-mass.

\begin{figure}
\includegraphics[width=8.5cm]{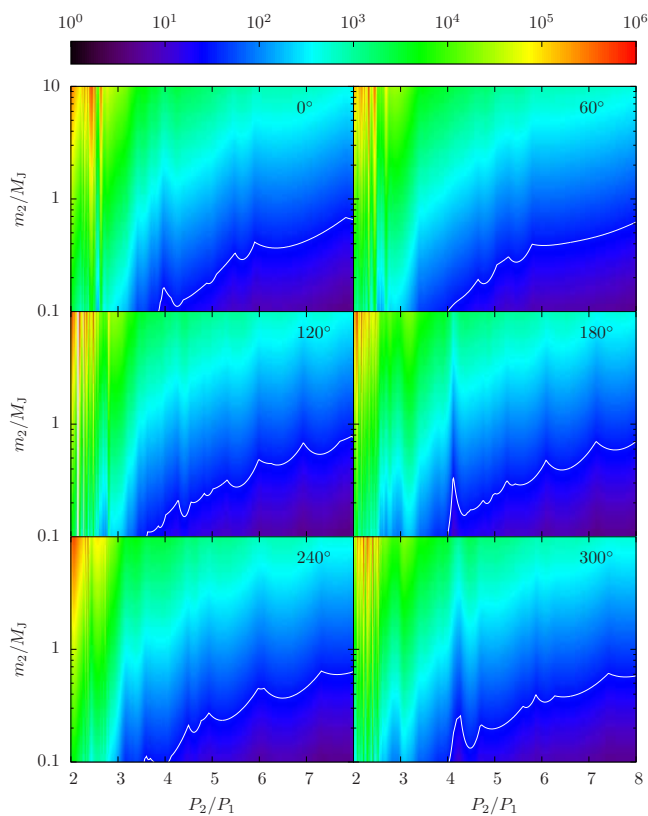}
\caption{Values of TTVs in the system of LHS 6343AC due to a hypothetical perturber.
The system is assumed to be circular and coplanar. Each panel shows TTVs for a different value
of the angular phase of the perturber. The units on the color scale are in seconds.  }
\label{fig14}
\end{figure}

\begin{figure}[t]
\centering
\includegraphics[trim=0mm 0mm 0mm 0mm, clip,width= 8.5 cm]{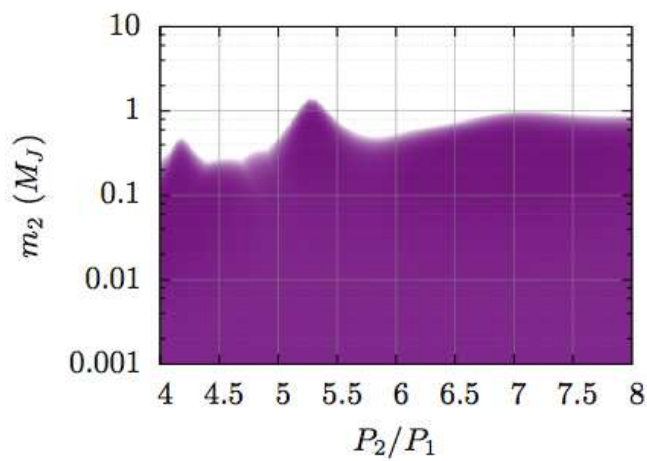}
\caption{Maximum mass of the perturber on circular orbit reproducing the
TTVs as in Figure 10, as a function of its orbital period.}
\label{fig:exslope}
\end{figure}

\section {Conclusion}

We presented a technique called the barycenter method for calculating the time of
the mid-transit in transit-timing studies. This method is based on the symmetry of the
light-curve, and has the advantage that is independent of the parameters of the system. In
other words, unlike other techniques for calculating the mid-transit timing, the results
obtained from the barycenter method will not change by changing the assumption on the parameters
of the central star. However, the fact that this method requires symmetry in the light-curve
implies that when the transit curve is not perfectly symmetric (i.e., when observational
points are missing, or because of starspots), large offsets may appear in the mid-transit timing measurements.
The application of the barycenter method to several known transiting systems showed that
the results obtained from this technique are comparable with those obtained from other methods. Our study indicates that for complete and symmetric transit lightcurves, the barycenter method achieves the same precision as in the model of \citet{Mandel-02}, but with fewer assumptions and much faster.

We used the barycenter method to calculate the times of mid-transits of the M+M binary star
LHS 6343. Our results indicated that as suggested by Johnson et al (2011),
the primary of this system is host to a smaller object with a period of $\sim 12.7$ days.
A study of the variations in the transit timing of this body (LHS 6343 C) points to the possibility that a small object with a mass no larger than 1 $M_{\rm J}$ may exist around LHS 6343 A, which can produce the O-C values lower than the upper values presented here. Whether such
an object actually exists requires more transit data and more observations of this system.

\begin{acknowledgements}

We acknowledge the support by the European Research Council/European Community under the
FP7 through Starting Grant agreement number 239953,
and by Funda\c{c}\~ao para a Ci\^encia e a Tecnologia (FCT) in
the form of grant reference PTDC/CTE-AST/098528/2008. NCS also acknowledge the support
from FCT through program Ci\^encia\,2007 funded by FCT/MCTES (Portugal) and
POPH/FSE (EC). G.B. thanks the Paris Observatory for providing the necessary computational resources for
this work. NH acknowledges support from the NASA/EXOB program through grant NNX09AN05G
and from the NASA Astrobiology Institute under
Cooperative Agreement NNA04CC08A at the Institute for Astronomy, University
of Hawaii.

\end{acknowledgements}

\bibliographystyle{aa}
\bibliography{mahlib}

\end{document}